\documentclass[aps,prb,twocolumn,groupedaddress,floatfix,amsmath,showpacs,lengthcheck]{revtex4-1}  

\usepackage{graphicx}
\usepackage{url}

\usepackage[colorlinks=true,%
urlcolor=blue,
linkcolor=blue,
citecolor=blue,
hypertexnames=true,
]{hyperref}

\begin{document}

\title{Spin Dependent Lifetimes and Spin-orbit Hybridization Points in Heusler Compounds}

\author{Steffen Kaltenborn}
\author{Hans Christian Schneider}
\email{hcsch@physik.uni-kl.de}
\affiliation{Physics Department and Research Center OPTIMAS, University of Kaiserslautern, 67663 Kaiserslautern, Germany}


\date{\today}

\begin{abstract}
We present an \emph{ab initio} calculation of the $\vec{k}$- and spin-resolved electronic lifetimes in the half-metallic Heusler compounds $\mathrm{Co}_2\mathrm{Mn}\mathrm{Si}$ and $\mathrm{Co}_2\mathrm{Fe}\mathrm{Si}$. We determine the spin-flip and spin-conserving contributions to the lifetimes and study in detail the behavior of the lifetimes around states that are strongly spin-mixed by spin-orbit coupling. We find that, for non-degenerate bands, the spin mixing alone does not determine the energy dependence of the (spin-flip) lifetimes. Qualitatively, the lifetimes reflect the lineup of electron and hole bands. We predict that different excitation conditions lead to drastically different spin-flip dynamics of excited electrons and may even give rise to an enhancement of the non-equilibrium spin polarization.
\end{abstract}

\maketitle

\section{Introduction}
Half metals with a high spin polarization at the Fermi energy, such as the so-called Heusler compounds, continue to be a focus of research because of their magneto-electronic properties, their suitability as materials for novel spintronics devices, and possibly their nontrivial topological properties.~\cite{ChadovNat,ChadovPRL} Much effort has gone into theoretically designing,~\cite{Balke,ChadovPRB3} fabricating,~\cite{Sakuraba,Mizukami1,Mizukami2} and characterizing~\cite{ChadovJMMM,Kallmayer,Fetzer,Ishida1,Ishida2} Heusler compounds with properties suitable for applications. An important aspect of these efforts continues to be the identification of compounds that display a full band gap in the minority channel. Because of the importance of correlation effects, state-of-the-art theoretical approaches have been refined and tested for these materials.~\cite{ChadovPRB1,ChadovPRB2}

Apart from the challenge of ab-initio description and design of half-metallic Heusler compounds, the problem also arises to understand carrier dynamics in an \emph{ideal} Heusler compound with a full gap,~\cite{Steil,Muenzenberg} for several reasons. 
First, the microscopic dynamics behind the macroscopic magnetization dynamics, which may show some interesting properties unique to half metals. For instance, it has been argued that the presence of a gap above the Fermi energy should have a significant impact on the demagnetization dynamics because of ``minority-state blocking'',~\cite{Muenzenberg} i.e., the unavailability of spin channels in the gap that would be available as final states for spin-flip scattering. Second, ``spin hot spots'' or spin-orbit hybridization points are generally believed to be of decisive importance for spin relaxation (or demagnetization) in semiconductors, metals,~\cite{Fabian,Mertig} and simple ferromagnets.~\cite{Mertig,Pickel,Koopmans} The band structure of half metals differs from those of metals by a lifted spin degeneracy and from those of simple ferromagnets by the existence of a band gap at the Fermi energy in one spin channel. These properties of the half-metallic band structure should also be reflected in the role played by spin hot-spots in their spin-dependent dynamics.

The present paper aims at investigating these two points for half-metallic ferromagnets. For two exemplary  Heusler compounds we compute \emph{ab initio} the spin-resolved electron and hole lifetimes due to carrier-carrier Coulomb interaction, and separate out the contributions of spin-flip and spin-conserving Coulomb scattering processes to the lifetimes. We can thus compare spin-flip and spin-conserving scattering processes for majority and minority electrons and holes. Our calculated lifetimes yield information on the carrier dynamics in the ab-initio ground-state band structure employed as input. A redistribution of carriers that goes beyond the dynamics described by the lifetimes (or out-scattering rates) is likely to change the quasi-particle states, and dynamics beyond carrier redistribution cannot, at present, be described in real bulk materials with complicated band structures. However, in our approach, one can still draw qualitative conclusions concerning, e.g., the available scattering phase space for minority and majority electrons also for elevated excitation conditions.

For the vast majority of $k$ points we find that the spin-conserving contributions to the lifetime are larger than the spin-flip contributions, both for majority and minority carriers. However, we find special $k$-points where the spin-flip contribution to the lifetime is \emph{larger} than the spin-conserving contribution.
We illustrate how this behavior may arise at $k$ points close to a minority band bottom.

\section{Density of states}
\begin{figure}
\includegraphics[width=0.48\textwidth]{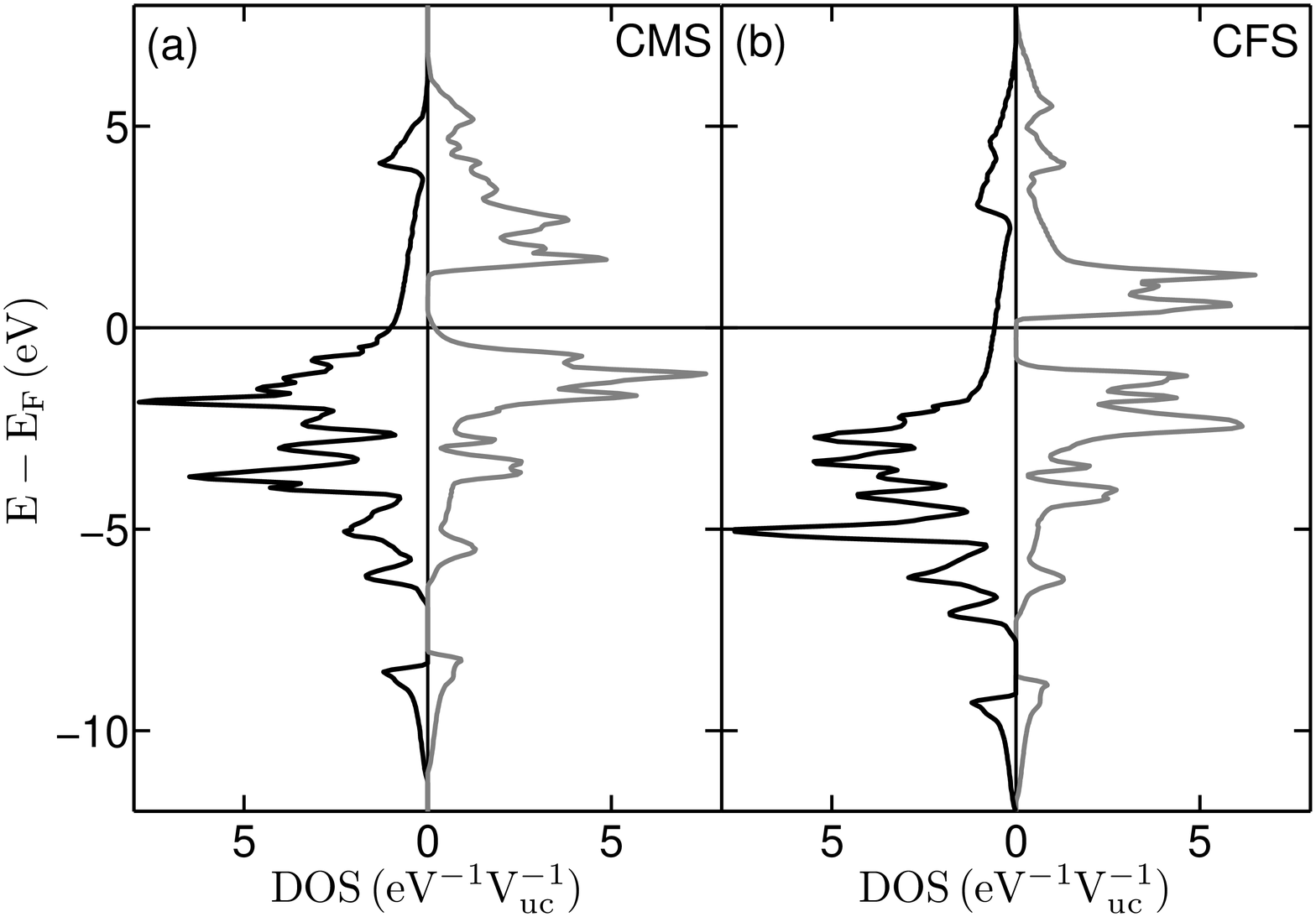}
\caption{\label{fig:dos}Electronic DOS per unit cell volume $V_\mathrm{uc}$ for CMS (a) and CFS (b). The black (grey) lines correspond to majority (minority) electrons. The calculation used $35^3$ $k$-points in the full 1.~Brillouin zone and included the first 60 (62) bands for the CMS (CFS) calculation.} 
\end{figure}

To begin with, we calculate the electronic density of states (DOS) for the two half-metallic Heusler compounds $\mathrm{Co}_2\mathrm{Mn}\mathrm{Si}$ (CMS) and $\mathrm{Co}_2\mathrm{Fe}\mathrm{Si}$ (CFS) within the full-potential linearized augmented plane wave (FP-LAPW) ELK code~\cite{ELK,Draxl} to reproduce earlier calculations.~\cite{Balke} The details concerning the crystal structure and the choice of the $U$ (LDA+$U$) parameter are given in  Ref.~\onlinecite{PRB13} and the resulting electronic DOS for the two Heusler compounds CMS and CFS is contained in Fig.~\ref{fig:dos}, which shows a good agreement with the DOS calculated in Refs.~\onlinecite{Balke,Wurmehl}. 

The difference in the DOS of the majority and minority electrons clearly shows the half-metallicity of the two compounds: The bands in the majority channel cross the Fermi energy, whereas a gap appears in the minority channel. The main difference between the two compounds is that the Fermi energy lies at the bottom of the band gap in CMS and at the upper end of the band gap in CFS. 

\section{Spin dependent lifetimes}
We calculate the $\vec{k}$- and band-resolved electronic lifetimes $\tau_{\vec{k}}^{\nu}$ due to the Coulomb interaction ab initio from standard expressions.~\cite{Zhukov,Ladstadter} These lifetimes are measured by 2-photon photoemission, and are sometimes referred to as inelastic lifetimes.~\cite{Aeschlimann-PRL10,Svenja} We briefly outline the calculational procedure here. Overlap matrix elements $B^{\mu\nu}_{\vec{k}\vec{q}}=\langle\psi_{\vec{k}+\vec{q}}^{\mu}|e^{i\vec{q}\cdot\vec{r}}|\psi_{\vec{k}}^{\nu}\rangle$ and the $\vec{k}$- and band-resolved energies $\epsilon_{\vec{k}}^{\mu}$ are calculated ab initio. Together with the distribution functions $f_{\vec{k}}^{\mu}$, this allows to calculate the dynamical, wave-vector dependent dielectric function $\varepsilon(\vec{q},\omega)$ in the random phase approximation (RPA) using our accurate wave-vector dependent linear tetrahedron method, which avoids the introduction of a broadening parameter.~\cite{PRB13} The computed complex dielectric functions were discussed in Ref.~\onlinecite{PRB13}, and are not reproduced here. We have checked that they agree with earlier calculations.~\cite{Picozzi,Meinert} The rates, i.e., the inverse lifetimes~$\gamma_{\vec{k}}^{\nu}=(\tau_{\vec{k}}^{\nu})^{-1}$ are determined by~\cite{Zhukov,Ladstadter}
\begin{equation}
\gamma_{\vec{k}}^{\nu}=\frac{2}{\hbar}\sum_{\mu\vec{q}}\frac{\Delta q^{3}}{(2\pi)^{3}}V_{q}\big|B_{\vec{k}\vec{q}}^{\mu\nu}\big|^{2}f^{\mu}_{\vec{k}+\vec{q}}\frac{\Im\varepsilon(\vec{q},\Delta E)}{|\varepsilon(\vec{q},\Delta E)|^{2}}
\label{eq:lifetime}
\end{equation}
for positive $\Delta E=\epsilon_{\vec{k}+\vec{q}}^{\mu}-\epsilon_{\vec{k}}^{\nu}$. For negative $\Delta E$, the distribution function has to be replaced by $-(1-f^{\mu}_{\vec{k}+\vec{q}})$. $\Delta q$ denotes the distance between two $\vec{q}$-points on the finite grid, and $V_q$ is the bare Coulomb potential. By representing the $\vec{k}$- and band-resolved wave-functions in a fixed spin basis,~\cite{Fabian,Koopmans} $|\psi_{\vec{k}}^{\mu}\rangle =a_{\vec{k}}^{\mu}\left|\uparrow\right\rangle +b_{\vec{k}}^{\mu}\left|\downarrow\right\rangle$, where the modulus of the coefficients $|a_{\vec{k}}^{\mu}|^2$ and $|b_{\vec{k}}^{\mu}|^2$ is extracted from the ELK-code,~\cite{ELK} we can also gain information about the lifetime behavior in the different spin channels.

\begin{figure}
\includegraphics[width=0.45\textwidth]{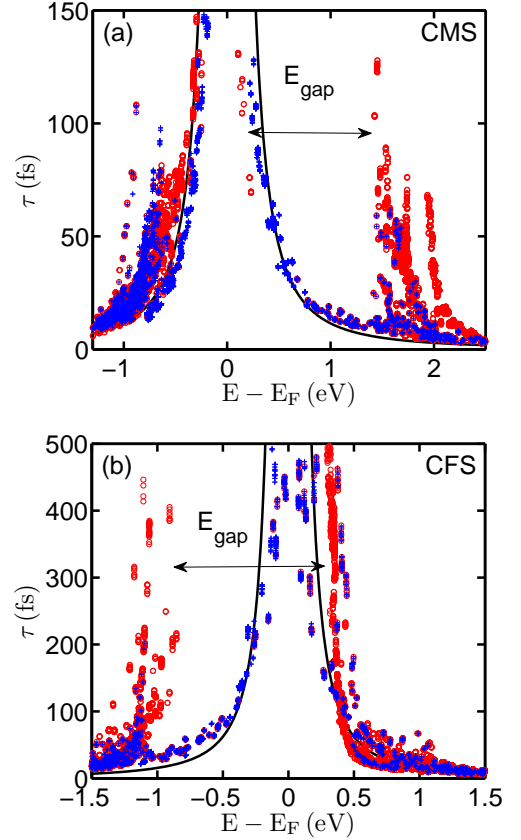}
\caption{\label{fig:lifetime}(Color online) Calculated energy- and spin-resolved majority ($+$) and minority ($\circ$) electronic lifetimes for the two Heusler compounds CMS (a) and CFS (b). For better readability the majority (minority) spin results are also color-coded in blue (red). The scatter results from different $\vec{k}$-points at the same energy. The gap energy corresponding to the DOS in Fig.~\ref{fig:dos} is indicated by the double arrow $E_{\mathrm{gap}}$ and the black solid lines correspond to a fit by an analytic Fermi liquid formula. The calculation used $13^3$ $\vec{k}$-points in the  full Brillouin zone.} 
\end{figure}

Figure~\ref{fig:lifetime} shows the calculated energy- and spin-resolved electronic lifetimes in CMS and CFS, which are determined from the band-dependent result by labeling all the states according to their dominating spin content. Due to the large number of bands in these compounds, there are many different $k$-points in the BZ with the same energy. The corresponding lifetimes therefore also show a wide range of values for a fixed energy, e.g.~at $1.2$--$0.5\,\mathrm{eV}$ ($1.5$--$1\,\mathrm{eV}$) below and $1.4$--$2.2\,\mathrm{eV}$ ($0.5$--$1.1\,\mathrm{eV}$) above the Fermi energy in the CMS (CFS) calculation. The spread in the calculated values marks the possible range of lifetimes for a energy-resolved measurement. 
In both materials, the \emph{majority} lifetimes near $E_{\mathrm{F}}$ increase following roughly a Fermi liquid behavior. From the analytic expression $\tau(E)=263r_{s}^{-5/2}(E-E_{\mathrm{F}})^{-2}$ (in units of fs and eV),~\cite{Chulkov, Ferrell, Quinn} the fit indicated by solid lines in Fig.~\ref{fig:lifetime} corresponds to the parameters $r_{s}(\mathrm{CMS})=3.5$ and $r_{s}(\mathrm{CFS})=3.1$. For the \emph{minority} spin, the lifetimes increase below or above the Fermi energy depending on the lineup of the band gap. In CMS, Fig.~\ref{fig:lifetime}(a), there is an increase at $1.4\,\mathrm{eV}$, which coincides with the upper end of the band gap in the minority channel. The same behavior can be found for CFS, Fig.~\ref{fig:lifetime}(b), where the lifetimes diverge around $-1\,\mathrm{eV}$, which corresponds to the bottom of the band gap in Fig.~\ref{fig:dos}(b). Thus, the DOS with the different locations of the minority band gaps is qualitatively reflected in the energy dependence of the minority lifetimes in the different materials. 

\begin{figure}
\includegraphics[width=0.45\textwidth]{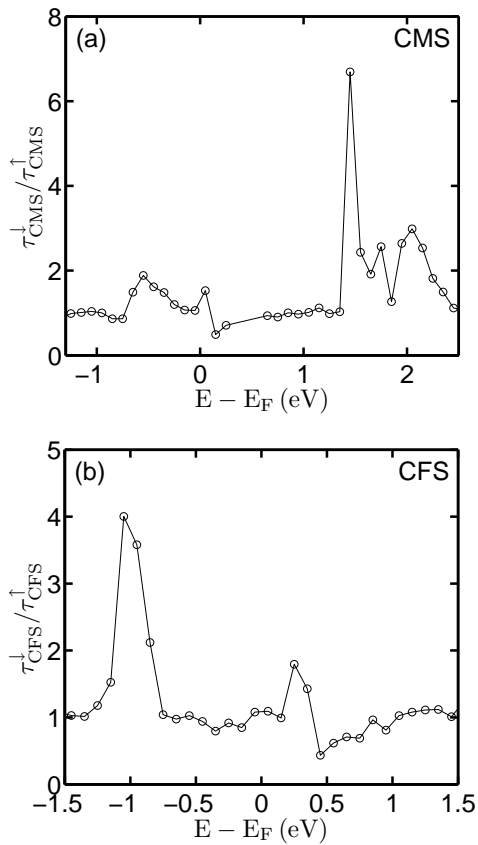}
\caption{\label{fig:lifetimeratio}Energy dependent ratio of averaged minority and majority electron lifetimes (``$\circ$''), $\tau^{\downarrow}/\tau^{\uparrow}$, for CMS (a) and CFS (b).} 
\end{figure}

To bring out the spin and energy dependence of the lifetimes more clearly, we average the lifetimes in Fig.~\ref{fig:lifetime} in bins of 100\,meV, which yields $\tau^\uparrow(E)$ and $\tau^\downarrow(E)$ and display the ratio $\tau^\uparrow/\tau^\downarrow$ of lifetimes for minority and majority electrons in Fig.~\ref{fig:lifetimeratio}. Note that, due to the spin-mixing, there are some $k$-points in the minority band gap, which are labeled as minority electrons, so that a ratio of minority and majority lifetimes in the band gap can be defined. Below $-0.8$\,eV and between 0 and 1.3\,eV the ratio $\tau^{\downarrow}/\tau^{\uparrow}$ for CMS (Fig.~\ref{fig:lifetimeratio}(a)) is around 1, i.e., there is no visible spin dependence. The most pronounced spin dependence is at the edges of the band gap, i.e., around 0 and 1.4\,eV. At the latter energy, the ratio of minority and majority electron lifetimes reaches a factor of around 7. Similar conclusions can be drawn for the ratio of minority and majority electron lifetimes in CFS, Fig.~\ref{fig:lifetimeratio}(b). Whereas no visible spin dependence occurs for a wide energy range, the ratio $\tau^{\downarrow}/\tau^{\uparrow}$ can increase to around 4 at the edges of the band gap. These differences may be of interest for applications in spintronics. Moreover, by using the lifetimes as input in transport calculations such as superdiffusive spin transport,~\cite{Battiato-PRBsuperdiff} these results can characterize the influence of transport on the (de)magnetization dynamics of the two Heusler compounds. 

\section{Spin-flip contribution to the lifetimes}

To extract information about the spin-dependent scattering pathways that contribute to the lifetimes at a given $k$ point, we separate spin-conserving (sc) and spin-flip (sf) contributions to the rates
$ (\tau_{\vec{k}}^{\nu})^{-1}=(\tau_{\vec{k},\mathrm{sc}}^{\nu})^{-1}+(\tau_{\vec{k},\mathrm{sf}}^{\nu})^{-1}$.
We achieve this by introducing the factors 
\begin{equation}
P_{\mathrm{sc}}^{(\nu,\vec{k})\to(\mu,\vec{k}+\vec{q})}=
|a_{\vec{k}}^{\nu}|^2 |a_{\vec{k}+\vec{q}}^{\mu}|^2  + |b_{\vec{k}+\vec{q}}^{\mu}|^2|b_{\vec{k}}^{\nu}|^2
\end{equation}
and $P_{\mathrm{sf}}=1-P_{\mathrm{sc}}$. For all individual transitions $(\nu,\vec{k})\to(\mu,\vec{k}+\vec{q})$ we take this as a measure for the change of up and down-spin components of the states. Formally, we  decompose the factors $|B_{\vec{k}\vec{q}}^{\mu\nu}|^{2}=(P_{\mathrm{sc}}+P_{\mathrm{sf}})|B_{\vec{k}\vec{q}}^{\mu\nu}|^{2}$ in the sum in Eq.~\eqref{eq:lifetime}. For comparatively small densities of excited carriers, these contributions to the lifetime determine whether a carrier preferentially flips its spin in a scattering event. As long as it is justified to work with lifetimes defined as in Eq.~\eqref{eq:lifetime}, the spin-flip and spin-conserving contributions to $\gamma$ are closely related to the spin-flip and spin-conserving dynamics of the excited carriers, and may be interpreted as the single-particle contribution to the magnetization dynamics.

\begin{figure}
\includegraphics[width=0.45\textwidth]{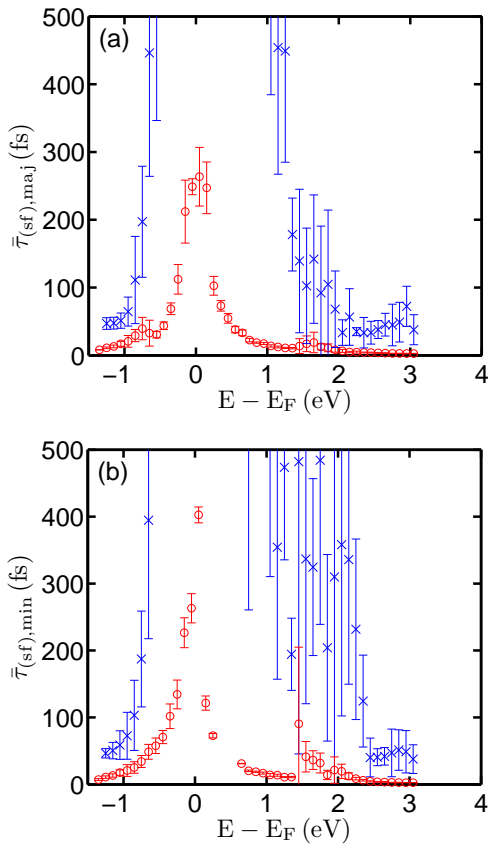}
\caption{\label{fig:spinlifetime}(Color online) Energetic average of spin-flip contributions $\bar{\tau}_{\vec{k},\mathrm{sf}}^{\nu}$ (blue ``$\times$'') and total (red ``$\circ$'') lifetimes $\bar{\tau}_{\vec{k}}^{\nu}$ in the majority (a) and the minority (b) spin channel of CMS. The error bars denote the standard deviation obtained from the scatter of the (spin-flip) lifetimes. We used $13^3$ $\vec{k}$-points in the full Brillouin zone.} 
\end{figure}

Figure~\ref{fig:spinlifetime} displays the energy-resolved total and spin-flip contribution to the lifetimes of CMS in the two spin channels. The energy resolved lifetimes $\bar{\tau}_{E(,\mathrm{sf})}$ are obtained by averaging over the $\vec{k}$ dependence of the rates and considering in each spin channel bins of $100\,\mathrm{meV}$. The calculated standard deviation therefore serves as an ``error bar'' due to the anisotropy of the crystal. Note that an increase (decrease) in the density of both majority electrons and minority holes leads to a increase (decrease) of the spin polarization. The reverse is true for the minority electrons and majority holes. The net spin flip results from a competition between these processes, and one can thus already draw some conclusions on the single-particle spin-flip dynamics after optical excitation from Fig.~\ref{fig:spinlifetime}. For instance, an excitation below the minority band gap, e.g., in the range $0.8$--$1.2\,\mathrm{eV}$, excites predominantly majority holes and electrons, for which the lifetimes are shown in Fig.~\ref{fig:spinlifetime}(a). The majority electrons produced by such an excitation, i.e., up to $1.2\,\mathrm{eV}$ above the Fermi energy have spin-flip lifetime contributions that are mostly larger than 500\,fs, which is the maximum shown in  Fig.~\ref{fig:spinlifetime}(a). For the holes in the range above $-1.2\,\mathrm{eV}$ the spin-flip lifetime contributions are much smaller than for the electrons, and around $-1$\,eV even on the order of the total lifetime. These optically excited majority holes flip their spin faster than the majority electrons, and our results predict an ultrafast increase of minority holes, which would contribute to an increase in spin polarization.

For larger excitation energies, around 2\,eV, say, one generally excites electrons and holes in both spin channels, and the spin-flip dynamics depend on the excitation conditions and are dominated by holes, similar to the case of 3d ferromagnets.~\cite{Sven-PRB} The generation of carriers by the optical excitation is, in turn, determined by the dipole matrix elements. Our results lend additional support to Ref.~\onlinecite{Steil}, where it was found that hole dynamics determine the spin polarization dynamics in CMS and CFS. In addition, the above results show that the effect of the ``blocking'' of transitions in the minority channel due to lack of minority states in the gap, as proposed in Ref.~\onlinecite{Muenzenberg}, is not sufficient to qualitatively determine the magnetization dynamics in half metals. Our results also show that, depending on excitation conditions, in the same material both an increase and a decrease of the spin polarization of excited carriers are possible. Such excitation dependent spin-flip dynamics should compete with super-diffusive spin-dependent transport, which was recently identified as the cause for different de/re-magnetization dynamics in ferromagnets.~\cite{Rudolf-NatComm}

\section{Lifetimes at spin-orbit hybridization points}
We next focus on particular states in the BZ in the vicinity of spin-orbit driven anticrossings (avoided crossings) of two bands with different spin orientations, the so-called spin-orbit hybridization points or spin hot-spots.~\cite{Mertig,Pickel,Fabian} For non-degenerate bands with opposite spin-mixing behavior around the hot-spot in the half metal CMS, we investigate in detail the different contributions to the lifetimes. Similar band properties in simple ferromagnets have been found in Refs.~\onlinecite{Feder,Pickel}. 

\begin{figure}
\includegraphics[width=0.45\textwidth]{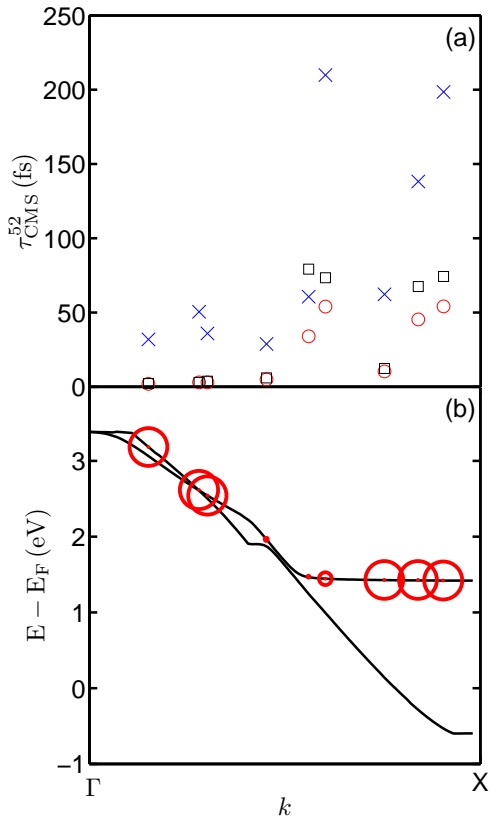}
\caption{\label{fig:spinhotspot}(Color online) (a) Momentum-resolved total lifetimes (red circles), spin-conserving (black squares) and spin-flip lifetime contributions (blue crosses) for electrons in CMS in the DFT-band 52 (minority band at the upper edge of the band gap). Note that all lifetimes for the three $k$ points at the right were compressed by a factor of 10. (b) Pair of minority (top) and majority (bottom) bands with opposite spin-mixing parameters. The diameter of the circles is proportional to $|b_{\vec{k}}^\mu|^2$; for the left and rightmost circles $|b|^2\simeq 1$.  We used $7^3$ and $13^3$ $\vec{k}$-points in the full BZ.} 
\end{figure}


In Fig.~\ref{fig:spinhotspot}(a) we plot $\vec{k}$-resolved total, spin-conserving and spin-flip lifetimes for a minority band (label ``$52$'') whose band bottom is at the top of the band gap of CMS. The corresponding band dispersion is the upper curve (with circles) in Fig.~\ref{fig:spinhotspot}(b). The $\vec{k}$ are in the $\Gamma$-$X$ direction. All three quantities increase for larger $k$, as this direction in $k$ space leads towards a minority band bottom, cf.~Fig.~\ref{fig:spinhotspot}(b). The most interesting behavior of the lifetimes in this particular band occurs at the $k$-value in the middle of the $\Gamma$-$X$ direction in Fig.~\ref{fig:spinhotspot}. The \emph{spin-flip} lifetime (blue cross) is below the \emph{spin-conserving} lifetime (black square), i.e., the spin-flip scattering probability is higher than that for spin-conserving transitions. This behavior of lifetimes is completely different from the vast majority of $k$ points, for which the spin-conserving contribution is orders of magnitude larger than the spin-flip one, and is not shared, in particular, by the majority band energetically directly below this minority band. This majority band, whose dispersion is also shown as the lower curve in Fig.~\ref{fig:spinhotspot}(b), has spin-conserving contributions that are at least five times larger than the spin-flip contribution, and never exhibits the ``inverted behavior''. 

The spin mixing parameter $|b^{\mu}_{\vec{k}}|^2$ for the minority band ``$52$'' is indicated by the diameter of the red circles in Fig.~\ref{fig:spinhotspot}(b) and shows that, going from the $\Gamma$ to the $X$ point, the direction of the average spin vector of the k-states is reversed, and then goes back to the direction identical to that at small $k$. The majority band below this minority band, whose dispersion is also shown in Fig.~\ref{fig:spinhotspot}(b) has exactly the opposite spin-mixing behavior compared to the minority band. This ``partner'' band is therefore mainly a majority band, only the spin mixing of the $k$ point in question is actually reversed.
This behavior shows that the spin mixing alone, which is the same (but opposite) for both bands, does not determine the ratio of spin-flip to spin-conserving scattering. Thus the spin-mixing alone does not uniquely identify the spin-flip dynamics close to a spin hot spot. It remains to be seen whether this special behavior may be exploited by selectively exciting minority electrons with $k$ vectors in this ($\Gamma$-$X$) direction. 
Note that we find many of these special $\vec{k}$-points in several bands below and above the Fermi energy for CMS and CFS, but this point is one of the few that lies on a main symmetry direction of the crystal.

\section{Conclusion}
In conclusion, we calculated spin-dependent lifetimes due to inelastic carrier Coulomb scattering for the two half-metallic Heusler compounds $\mathrm{Co}_2\mathrm{Mn}\mathrm{Si}$ and $\mathrm{Co}_2\mathrm{Fe}\mathrm{Si}$. We showed that the majority bands crossing the Fermi energy exhibit a Fermi liquid behavior for the lifetimes. The electronic lifetimes in minority bands increase towards the respective band bottoms bordering on the band gap. We argued that the single-particle contribution to the laser-excited spin-flip dynamics of these compounds strongly depends on the excitation process, with holes generally dominating the spin-flip scattering dynamics. This should result in an \emph{increase}  in spin polarization for excitation with photon energies below 1.2\,eV. We also showed the existence of a majority-minority pair of bands with complementary spin mixing behavior, but very different spin-flip and spin-conserving contributions to the lifetime. The (modulus of the) spin-mixing parameter $b^2$ therefore does not uniquely identify the spin-flip scattering dynamics in a nondegenerate Bloch state.

\begin{acknowledgments}
We acknowledge a CPU-time grant at the J\"{u}lich Supercomputer Center (JSC).
\end{acknowledgments}

\end{document}